\documentclass[conference]{IEEEtran}

\usepackage{amsmath}
\usepackage{amssymb}
\usepackage[dvips]{graphicx}
\usepackage{epsfig}

\newcommand{\figsize}{0.4}

\newcommand{\tsnr}{{\text{\footnotesize{SNR}}}}

\newtheorem{prop:asympcap}{Theorem}
\newtheorem{prop:flashminbitenergy}[prop:asympcap]{Theorem}
\newtheorem{prop:flashbitenergy}[prop:asympcap]{Theorem}
\newtheorem{prop:pasympcap}[prop:asympcap]{Theorem}
\newtheorem{lemma:bitenergylow}[prop:asympcap]{Theorem}

\begin{document}


\title{To Cooperate, or Not to Cooperate in Imperfectly-Known Fading Channels}



%
\author{\authorblockN{Junwei Zhang \quad Mustafa Cenk Gursoy}
\authorblockA{Department of Electrical Engineering\\
University of Nebraska-Lincoln, Lincoln, NE 68588\\ Email:
jzhang13@bigred.unl.edu, gursoy@engr.unl.edu}}


\maketitle

\begin{abstract}\footnote{This work was supported in part by the NSF CAREER Grant CCF-0546384}
In this paper, communication over imperfectly-known fading channels
with different degrees of cooperation is studied. The three-node
relay channel is considered. It is assumed that communication starts
with the network training phase in which the receivers estimate the
fading coefficients of their respective channels. In the data
transmission phase, amplify-and-forward and decode-and-forward
relaying schemes are employed. For different cooperation protocols,
achievable rate expressions are obtained. These achievable rate
expressions are then used to find the optimal resource allocation
strategies. In particular, the fraction of total time or bandwidth
that needs to be allocated to the relay for best performance is
identified. Under a total power constraint, optimal allocation of
power between the source and relay is investigated. Finally, bit
energy requirements in the low-power regime are studied.
\end{abstract}
\section{Introduction}
In wireless communications, deterioration in  performance is
experienced due to various impediments such as interference,
fluctuations in power due to reflections and attenuation, and
randomly-varying channel conditions caused by mobility and changing
environment. Recently, cooperative wireless communications has
attracted much interest as a technique that can mitigate these
degradations and provide higher rates or improve the reliability
through diversity gains. The relay channel was first introduced by
van der Meulen in \cite{meulen}, and initial research is primarily
conducted to understand the rates achieved in relay channels
\cite{cover}, \cite{Gamal}. More recently, diversity gains of
cooperative transmission techniques have been studied in~\cite{jnl}
and \cite{jnlbook} where several two-user cooperative protocols have
been proposed, with amplify-and-forward~(AF) and
decode-and-forward~(DF) being the two basic modes. In~\cite{Nabar},
three different time-division AF and DF cooperative protocols  with
different the degrees of broadcasting and receive collision are
studied. Most work in this area has assumed that the channel
conditions are perfectly known at the receiver and/or transmitter
sides. However, especially in mobile applications in which the
channel changes randomly, the channel conditions can only be learned
imperfectly.

In this paper, we study the impact of cooperation in
imperfectly-known fading channels where a priori unknown fading
coefficients are estimated at the receivers with the assistance of
pilot symbols. We obtain achievable rates for AF and DF relaying
techniques. Achievable rates are subsequently used to find the
optimal fraction of total time or bandwidth allocated to the relay,
enabling us to identify the optimal degree of cooperation. Under
total power constraints, optimal power allocation strategies are
determined. Moreover, we investigate the energy efficiency in the
low-power regime.

\section{Channel Model} \label{sec:model}

We consider the three-node relay network which consists of a source,
destination, and a relay node. Source-destination, source-relay, and
relay-destination channels are modeled as  Rayleigh block-fading
channels with fading coefficients denoted by $h_{sr}$, $ h_{sd}$,
and $h_{rd}$, respectively for each channel. Due to the block-fading
assumption, the fading coefficients $h_{sr}\sim\mathcal {C}\mathcal
{N}(0,{\sigma_{sr}}^2)$, $h_{sd}\sim\mathcal {C}\mathcal
{N}(0,{\sigma_{sd}}^2)$, and $h_{rd}\sim\mathcal {C}\mathcal
{N}(0,{\sigma_{rd}}^2)$ \footnote{$x\sim\mathcal {C}\mathcal
{N}(d,{\sigma^2)}$ is used to denote a proper complex Gaussian
random variable with mean $d$ and variance $\sigma^2$.} stay
constant for a block of $m$ symbols before they assume independent
realizations for the following block. It is assumed that the source,
relay, and destination nodes do not have prior knowledge of the
instantaneous realizations of the fading coefficients. Hence, the
transmission is conducted in two phases: network training phase in
which the fading coefficients are estimated at the receivers, and
data transmission phase. Overall, the source and relay are subject
to the following power constraints in one block:
$\|{\mathbf{x}_{s,t}}\|^2+E\{\|{\mathbf{x}_s}\|^2\}\leq mP_{s}$ and
$\|{\mathbf{x}_{r,t}}\|^2+E\{\|{\mathbf{x}_r}\|^2\}\leq mP_{r}$
where $\mathbf{x}_{s,t}$ and $\mathbf{x}_{r,t}$ are the source and
relay training signal vectors, respectively, and $\mathbf{x}_{s}$
and $\mathbf{x}_{r}$ are the corresponding source and relay data
vectors.
\subsection{Network Training Phase}
Each block transmission starts with the training phase. In the first
symbol period, source transmits a pilot symbol to enable the relay
and destination to estimate channel coefficients $h_{sr}$ and
$h_{sd}$. In the average power limited case, sending a single pilot
is optimal because instead of increasing the number of pilot
symbols, a single pilot with higher power can be used. The signals
received by the relay and destination, respectively, are
$y_{r,t}=h_{sr}x_{s,t}+n_r$ and $y_{d,t}=h_{sd}x_{s,t}+n_d$.
Similarly, in the second symbol period, relay transmits a pilot
symbol to enable the destination to estimate the channel coefficient
$h_{rd}$. The signal received by the destination is
$y_{d,t}^{r}=h_{rd}x_{r,t}+n_d^r$. In the above formulations,
$n_r\sim\mathcal {C}\mathcal {N}(0,N_0)$, $n_d\sim\mathcal
{C}\mathcal {N}(0,N_0)$, and $n_d^r\sim\mathcal {C}\mathcal
{N}(0,N_0)$ represent independent Gaussian noise samples at the
relay  and the destination nodes.

In the training process, it is assumed that the receivers employ
minimum mean-square error (MMSE) estimation. Let us assume that the
source allocates $\delta_s$ of its total power for training while
the relay allocates $\delta_r$ of its total power for training. As
described in  \cite{gursoy}, the MMSE estimate of $h_{sr}$ is given
by
$\hat{h}_{sr}=\frac{\sigma_{sr}^2\sqrt{\delta_smP_s}}{\sigma_{sr}^2\delta_smP_s+N_0}y_{r,t}$
where $y_{r,t}\sim\mathcal {C}\mathcal
{N}(0,\sigma_{sr}^2\delta_smP_s+N_0)$. We denoted by
 $\tilde{h}_{sr}$  the estimate error which is a zero-mean complex
Gaussian random variable with variance
$var(\tilde{h}_{sr})=\frac{\sigma_{sr}^2N_0}{\sigma_{sr}^2\delta_smP_s+N_0}$.
Note that we can write $h_{sr}=\hat{h}_{sr} +\tilde{h}_{sr}$.
Similar expressions are obtained for
$\hat{h}_{sd},\tilde{h}_{sd},\hat{h}_{rd}$, and $\tilde{h}_{rd}$
which are the estimates and estimate errors of the fading
coefficients $h_{sd}$ and $h_{rd}$, respectively.
%

\subsection {Data Transmission Phase}

The practical relay node usually cannot transmit and receive data
simultaneously. Thus, we assume that the relay works under
half-duplex constraint. As discussed in the previous section, within
a block of $m$ symbols, the first two symbols are allocated for
channel training. In the remaining duration of $m-2$ symbols, data
transmission takes place. We introduce the relay transmission
parameter $\alpha$ and assume that $\alpha(m-2)$ symbols are
allocated for relay transmission. Hence, $\alpha$ can be seen as the
fraction of total time or bandwidth allocated to the relay. Note
that the parameter $\alpha$ enables us to control the degree of
cooperation. In this paper, we consider three relaying schemes:
Amplify and Forward, Decode and Forward with repetition channel
coding, and Decode and Forward with parallel channel coding.

\subsubsection{AF and repetition DF} In AF and repetition DF, since the relay either amplifies the
received signal, or decodes it but uses the same codebook as the
source to forward the data, cooperative transmission takes place in
the duration of $2\alpha(m-2)$ symbols. The remaining duration of
$(1-2\alpha)(m-2)$ symbols is allocated to unaided direct
transmission from the source to the destination. It is obvious that
we have $0<\alpha \leq 1/2$ in this setting. Therefore, $\alpha =
1/2$ models full cooperation while we have noncooperative
communications as $\alpha \to 0$.

In these protocols, the input-output relations are expressed as
follows: $\mathbf{y}_{d1}=h_{sd}\mathbf{x}_{s1}+\mathbf{n}_d$,
$\mathbf{y}_r=h_{sr}\mathbf{x}_{s21}+\mathbf{n}_r$,
$\mathbf{y}_{d2}=h_{sd}\mathbf{x}_{s21}+\mathbf{n}_d$,
$\mathbf{y}_d^r=h_{sd}\mathbf{x}_{s22}+h_{rd}\mathbf{x}_{r}+\mathbf{n}_d^r$.
Above, $\mathbf{x}_{s1},\mathbf{x}_{s21},\mathbf{x}_{s22}$, which
have respective dimensions $(1-2\alpha)(m-2)$, $\alpha(m-2)$ and
$\alpha(m-2)$, represent the source data vectors sent in direct
transmission, cooperative transmission when relay is listening and
cooperative transmission when relay is transmitting, respectively.
Note that we assume in this case that the source transmits all the
time. $\mathbf{x}_{r}$ is the relay's data vector with dimension
$\alpha(m-2)$. $\mathbf{y}_{d1},\mathbf{y}_{d2},\mathbf{y}_d^r$ are
the corresponding received vectors at the destination, and
$\mathbf{y}_r$ is the received vector at the relay. The input vector
$\mathbf{x}_s$ now is defined as
$\mathbf{x}_{s}=[\mathbf{x}_{s1}^T,\mathbf{x}_{s21}^T,\mathbf{x}_{s22}^T]^T$,
and $\mathbf{x}_s$ and $\mathbf{x}_{r}$ vectors are assumed to be
composed of independent random variables with equal energy. Hence,
the corresponding covariance matrices are
\begin{equation}\label{xsp1}
E\{\mathbf{x}_{s}\mathbf{x}_{s}^\dagger\}=P_{s}'\,\,\mathbf{I}=\frac{(1-\delta_s)mP_s}{m-2}\,\,\mathbf{I},
\end{equation}
\begin{equation}\label{xrp2}
E\{\mathbf{x}_{r}\mathbf{x}_{r}^\dagger\}=P_{r}'\,\,\mathbf{I}=\frac{(1-\delta_r)mP_r}{(m-2)\alpha}\,\,\mathbf{I}.
\end{equation}
\subsubsection{Parallel DF}
In DF with parallel channel coding, we simplify the transmission by
assuming that the source becomes silent while relay is transmitting
information. Thus, source transmits over a duration of
$(1-\alpha)(m-2)$ symbols. Now, the range of $\alpha$ is $0<
\alpha<1$. In this case, the input-output relations are given by:
$\mathbf{y}_r=h_{sr}\mathbf{x}_{s}+\mathbf{n}_r$,
$\mathbf{y}_{d}=h_{sd}\mathbf{x}_{s}+\mathbf{n}_d$,
$\mathbf{y}_d^r=h_{rd}\mathbf{x}_{r}+\mathbf{n}_d^r$. The dimensions
of the vectors $\mathbf{x}_{s}, \mathbf{y}_{d}, \mathbf{y}_r$ are
$(1-\alpha)(m-2)$, while $\mathbf{x}_{r},\mathbf{y}_d^r$ are vectors
of dimension $\alpha(m-2)$. In this case, the covariance matrix for
relay data vector remains the same in as (\ref{xrp2}) while the
covariance for source data vector $\mathbf{x}_{s}$ is
\begin{equation}\label{xsp2}
E\{\mathbf{x}_{s}\mathbf{x}_{s}^\dagger\}=P_{s1}'\,\,\mathbf{I}=\frac{(1-\delta_s)mP_s}{(m-2)(1-\alpha)}\,\,\mathbf{I}.
\end{equation}

\section{Achievable Rates}
In this section, we find achievable rate expressions for the three
relaying protocols described in Section \ref{sec:model}.
Using the same techniques described in \cite{Zhagur}, we can show
that capacity lower bounds are obtained when the channel estimation
error is assumed to be another source of Gaussian noise. This is due
to the fact that Gaussian noise is the worst uncorrelated noise for
the Gaussian model. The achievable rate expressions are provided
below. The proofs are omitted due to lack of space.
\begin{prop:asympcap} \label{prop:overlapped}
An achievable rate expression  for AF transmission scheme is given
by
\begin{align}\label{cafover1}
I_{low} =& \frac{1}{m}E\Bigg\{
\Bigg[(1-2\alpha)(m-2)\log(1+\frac{P_{s}'|\hat{h}_{sd}|^2}{\sigma_{z_{d}}^2})+(m-2)\alpha \nonumber\\
&\log\Big(1+\frac{P_{s}'|\hat{h}_{sd}|^2}{\sigma_{z_{d}}^2}+f\bigg(\frac{P_{s}'|\hat{h}_{sr}|^2}{\sigma_{z_{r}}^2},\frac{P_{r}'|\hat{h}_{rd}|^2}{\sigma_{z_{d}^{r}}^2}\bigg)
\nonumber\\
&+q\bigg(\frac{P_{s}'|\hat{h}_{sd}|^2}{\sigma_{z_{d}}^2},\frac{P_{s}'|\hat{h}_{sd}|^2}{\sigma_{z_{d}^{r}}^2},\frac{P_{s}'|\hat{h}_{sr}|^2}{\sigma_{z_{r}}^2},\frac{P_{r}'|\hat{h}_{rd}|^2}{\sigma_{z_{d}^{r}}^2}\bigg)
\Big) \Bigg]\Bigg\}
\end{align}
where $f(.)$, $q(.)$ are defined as $f(x,y)=\frac{xy}{1+x+y}$
,$q(a,b,c,d)=\frac{(1+a)b(1+c)}{1+c+d}$. Moreover {\small
\begin{equation}\label{pssdzd2}
\frac{P_{s}'|\hat{h}_{sd}|^2}{\sigma_{z_{d}}^2}=\frac{\delta_s(1-\delta_s)m^2P_s^2\sigma_{sd}^4}{(1-\delta_s)mP_s\sigma_{sd}^2N_0+(m-2)(\sigma_{sd}^2\delta_smP_s+N_0)N_0}|w_{sd}|^2
\end{equation}
\begin{equation}\label{pssrzr2}
\frac{P_{s}'|\hat{h}_{sr}|^2}{\sigma_{z_{r}}^2}=\frac{\delta_s(1-\delta_s)m^2P_s^2\sigma_{sr}^4}{(1-\delta_s)mP_s\sigma_{sr}^2N_0+(m-2)(\sigma_{sr}^2\delta_smP_s+N_0)N_0}|w_{sr}|^2
\end{equation}
\begin{eqnarray}\label{pssdzdr2}
\frac{P_{s}'|\hat{h}_{sd}|^2}{\sigma_{z_{d}^{r}}^2}=\frac{\delta_s(1-\delta_s)m^2P_s^2\sigma_{sd}^4(\sigma_{rd}^2\delta_rmP_r+N_0)}{(m-2)(\sigma_{sd}^2\delta_smP_s+N_0)(\sigma_{rd}^2\delta_rmP_r+N_0)N_0}\nonumber\\
\frac{}{+(1-\delta_r)mP_r\sigma_{rd}^2N_0(\sigma_{sd}^2\delta_smP_s+N_0)/ \alpha}\nonumber\\
\frac{}{+(1-\delta_s)mP_s\sigma_{sd}^2N_0(\sigma_{rd}^2\delta_rmP_r+N_0)}|w_{sd}|^2
\end{eqnarray}
\begin{eqnarray}\label{prrdzdr2}
\frac{P_{r}'|\hat{h}_{rd}|^2}{\sigma_{z_{d}^{r}}^2}=\frac{\delta_r(1-\delta_r)m^2P_r^2\sigma_{rd}^4(\sigma_{sd}^2\delta_smP_s+N_0)/\alpha}{(m-2)(\sigma_{sd}^2\delta_smP_s+N_0)(\sigma_{rd}^2\delta_rmP_r+N_0)N_0}\nonumber\\
\frac{}{+(1-\delta_r)mP_r\sigma_{rd}^2N_0(\sigma_{sd}^2\delta_smP_s+N_0)/\alpha}\nonumber\\
\frac{}{+(1-\delta_s)mP_s\sigma_{sd}^2N_0(\sigma_{rd}^2\delta_rmP_r+N_0)}|w_{rd}|^2
\end{eqnarray}
} and here and henceforth $w_{sr},w_{sd}, w_{rd}$ are i.i.d with
distribution $\mathcal {C}\mathcal {N}(0,1)$.
\end{prop:asympcap}
\begin{prop:asympcap} \label{prop:DFRover}
An achievable rate expression for DF with repetition channel coding
transmission scheme is given by {\footnotesize
\begin{equation}\label{DFC1}
I_{low}=\frac{1}{m} \Big\{
(1-2\alpha)(m-2)E_{w_{sd}}\log(1+\frac{P_{s}'|\hat{h}_{sd}|^2}{\sigma_{z_{d}}^2})
+(m-2)\alpha \min\{I_1,I_2\}\Big\}
\end{equation}
where
\begin{equation}\label{DFC11}
I_1=E_{w_{sr}}\Bigg[\log\Bigg(1+\frac{P_{s}'|\hat{h}_{sr}|^2}{\sigma_{z_{r}}^2}\Bigg)\Bigg],
\end{equation}
\begin{eqnarray}
I_2=E_{w_{sd}}E_{w_{rd}}\Bigg[\log
\Big(1+\frac{P_{s}'|\hat{h}_{sd}|^2}{\sigma_{z_{d}}^2}+\frac{P_{r}'|\hat{h}_{rd}|^2}{\sigma_{z_{d}^r}^2}
\nonumber
\\+\frac{P_{s}'|\hat{h}_{sd}|^2}{\sigma_{z_{d}^r}^2}+
\frac{P_{s}'|\hat{h}_{sd}|^2}{\sigma_{z_{d}}^2}\frac{P_{s}'|\hat{h}_{sd}|^2}{\sigma_{z_{d}^r}^2}\Big)\Bigg]
\end{eqnarray}}
where $\frac{P_{s}'|\hat{h}_{sd}|^2}{\sigma_{z_{d}}^2}$,
$\frac{P_{s}'|\hat{h}_{sr}|^2}{\sigma_{z_{r}}^2}$,$\frac{P_{s}'|\hat{h}_{sd}|^2}{\sigma_{z_{d}^r}^2}$,
$\frac{P_{r}'|\hat{h}_{rd}|^2}{\sigma_{z_{d}^r}^2}$ are the same as
in (\ref{pssdzd2})-(\ref{prrdzdr2})
\end{prop:asympcap}

With regard to parallel DF, based on \cite{liang}, we can write the
capacity lower bound as
\begin{align*}
I_{low}=\sup_{p_{x_{s}}(\cdot),p_{x_{r}}(\cdot )}\min \{ &(1-\alpha)
\emph{I}(\mathbf{x}_s ;\mathbf{y}_r|\hat{h}_{sr}),(1-\alpha)
\emph{I}(\mathbf{x}_s ;\mathbf{y}_d|\hat{h}_{sd})\\
\nonumber &+\alpha \emph{I}(\mathbf{x}_r
;\mathbf{y}_d^r|\hat{h}_{rd})\}.
\end{align*}
Using similar methods as before, we obtain the following result.

\begin{prop:asympcap} \label{prop:asympcap}
An achievable rate of DF with parallel channel coding scheme is
given by
\begin{align}\label{DFP}
I_{low}=\min \Bigg
\{E\left\{\frac{(1-\alpha)(m-2)}{m}\log\Big(1+\frac{P_{s1}'|\hat{h}_{sr}|^2}{\sigma_{z_{r1}}^2}\Big)\right\},\nonumber\\
\nonumber
E\left\{\frac{(1-\alpha)(m-2)}{m}\log\Big(1+\frac{P_{s1}'|\hat{h}_{sd}|^2}{\sigma_{z_{d1}}^2}\Big)\right.+\\
\nonumber \left.\frac{\alpha (m-2)}{m}\log
\Big(1+\frac{P_{r1}'|\hat{h}_{rd}|^2}{\sigma_{z_{d1}^r}^2}\Big)
\Bigg\}\right\}
\end{align}
where {\footnotesize

\begin{equation*}
\frac{P_{s1}'|\hat{h}_{sd}|^2}{\sigma_{z_{d1}}^2}=\frac{\delta_s(1-\delta_s)m^2P_s^2\sigma_{sd}^4/(1-\alpha)|w_{sd}|^2}{(1-\delta_s)mP_s\sigma_{sd}^2N_0/(1-\alpha)+(m-2)(\sigma_{sd}^2\delta_smP_s+N_0)N_0}
\end{equation*}

\begin{equation*}
\frac{P_{s1}'|\hat{h}_{sr}|^2}{\sigma_{z_{r1}}^2}=\frac{\delta_s(1-\delta_s)m^2P_s^2\sigma_{sr}^4/(1-\alpha)|w_{sr}|^2}{(1-\delta_s)mP_s\sigma_{sr}^2N_0/(1-\alpha)+(m-2)(\sigma_{sr}^2\delta_smP_s+N_0)N_0}
\end{equation*}

\begin{equation*}
\frac{P_{r1}'|\hat{h}_{rd}|^2}{\sigma_{z_{d1}^r}^2}=\frac{\delta_r(1-\delta_r)m^2P_r^2\sigma_{rd}^4/\alpha|w_{rd}|^2}{(1-\delta_r)mP_r\sigma_{rd}^2N_0/\alpha+(m-2)(\sigma_{rd}^2\delta_smP_r+N_0)N_0}
\end{equation*}
}
\end{prop:asympcap}
\section{Numerical Results}
\begin{figure}
\begin{center}
\includegraphics[width = \figsize\textwidth]{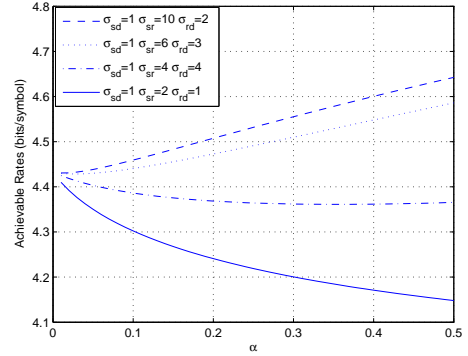}
\caption{  AF achievable rate  vs. $\alpha$ when $P_s=P_r=50,
\delta_s=\delta_r=0.1$} \label{fig:overAF1}
\end{center}
\end{figure}

\begin{figure}
\begin{center}
\includegraphics[width = \figsize\textwidth]{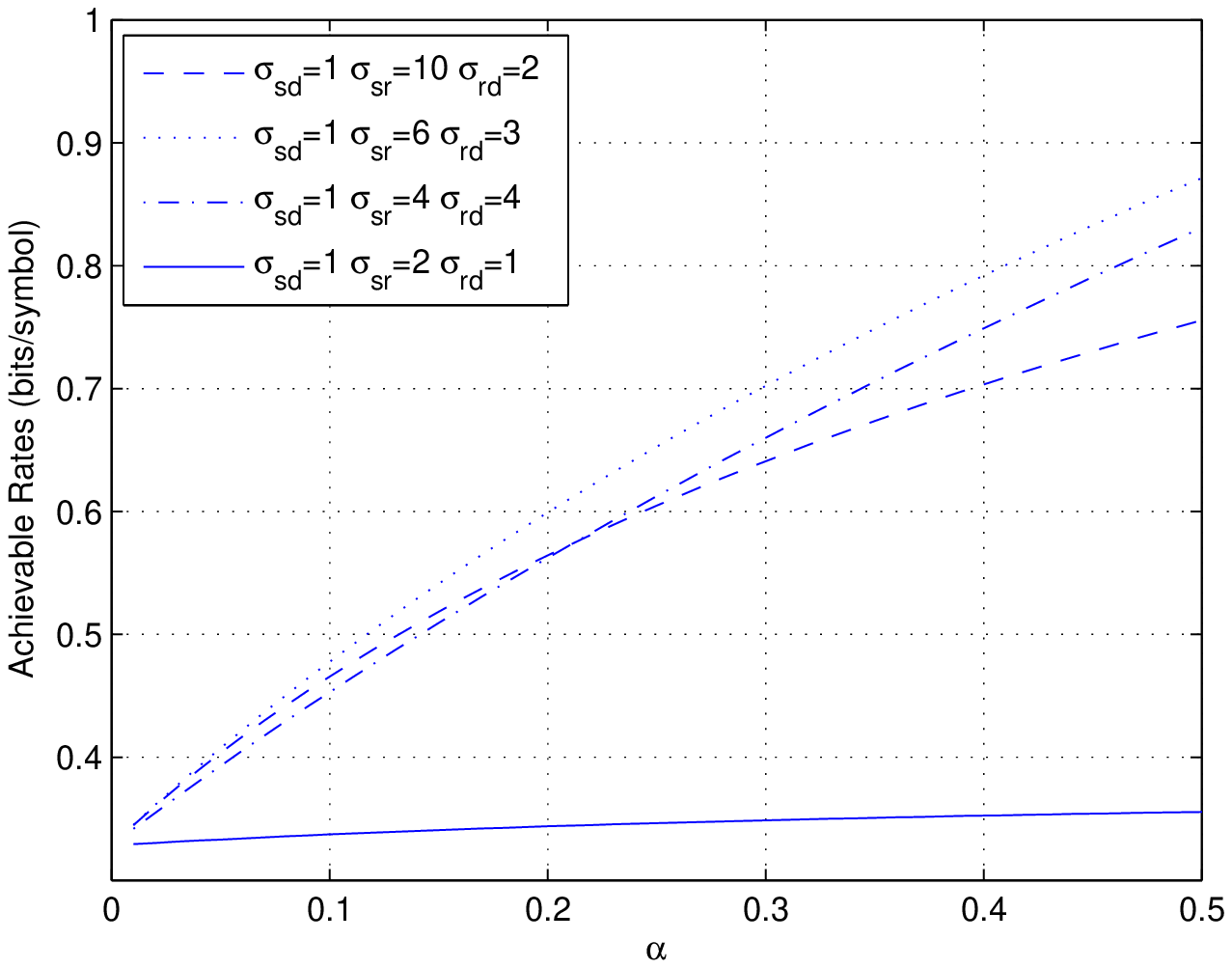}
\caption{  AF achievable rate  vs. $\alpha$ when $P_s=P_r=0.5,
\delta_s=\delta_r=0.1$} \label{fig:overAF2}
\end{center}
\end{figure}

\begin{figure}
\begin{center}
\includegraphics[width = \figsize\textwidth]{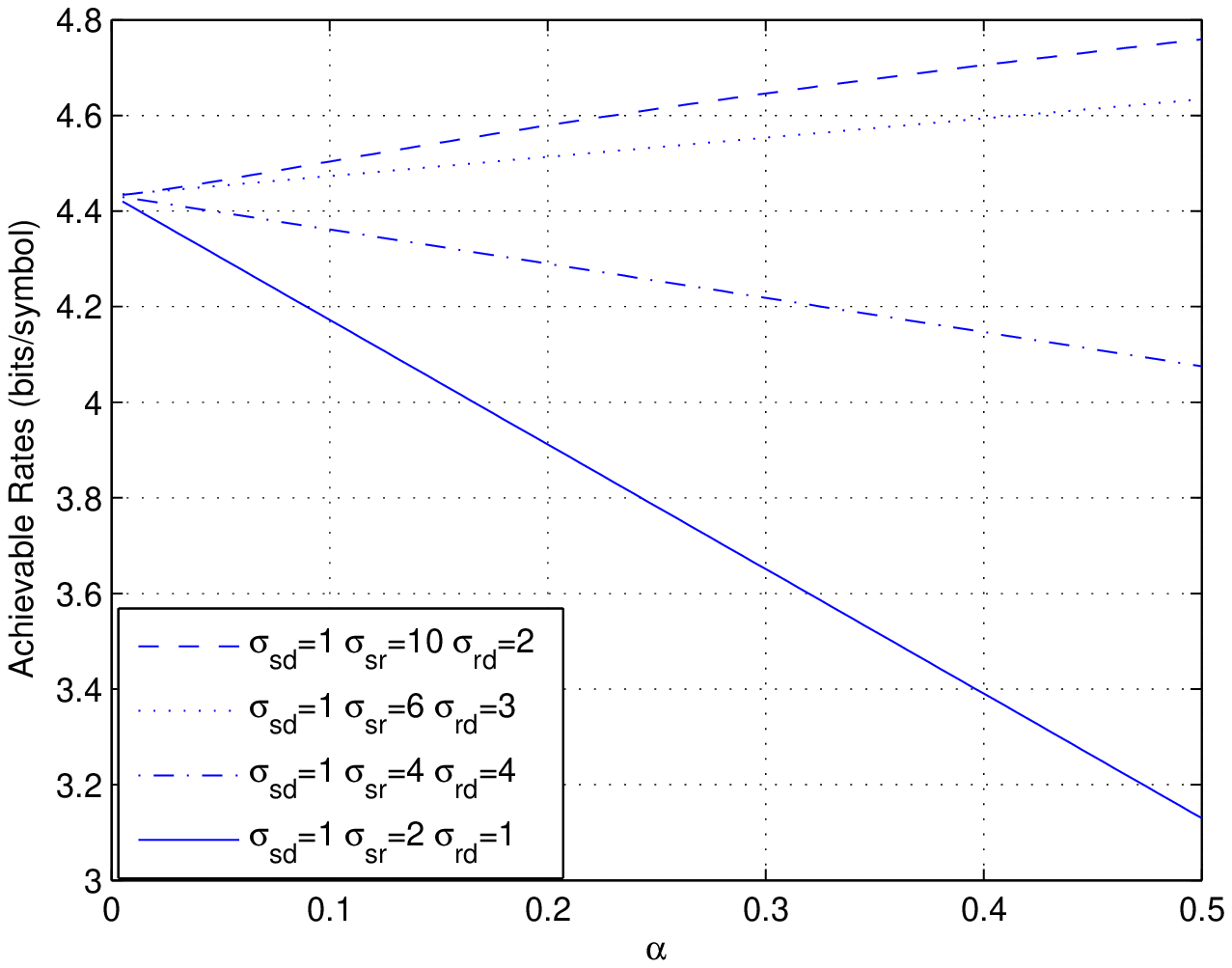}
\caption{  DF with repetition coding achievable rate  vs. $\alpha$
when $P_s=P_r=50, \delta_s=\delta_r=0.1$} \label{fig:overDF}
\end{center}
\end{figure}

\begin{figure}
\begin{center}
\includegraphics[width = \figsize\textwidth]{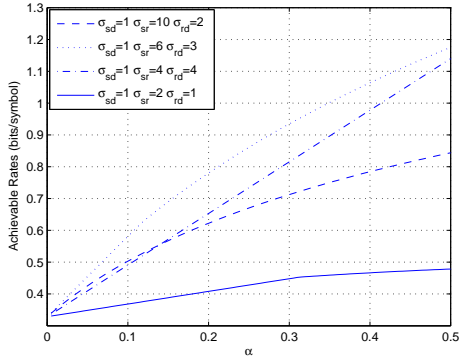}
\caption{  DF with repetition coding achievable rate  vs. $\alpha$
when $P_s=P_r=0.5, \delta_s=\delta_r=0.1$} \label{fig:overDF2}
\end{center}
\end{figure}

We first analyze the effect of the degree of cooperation on the
performance in AF and repetition DF. Figures
\ref{fig:overAF1}-\ref{fig:overDF2} plot the achievable rates as a
function of $\alpha$. Achievable rates are obtained for different
channel qualities given by the standard deviations $\sigma_{sd},
\sigma_{sr},$ and $\sigma_{rd}$ of the fading coefficients.
We observe that if the input power is high, $\alpha$ should be
either $0.5$ or close to zero depending on the channel qualities. On
the other hand, $\alpha=0.5$ always gives us the best performance at
low $\tsnr$ levels regardless of the channel qualities. Hence, while
cooperation is beneficial in the low-$\tsnr$ regime, noncooperative
transmissions might be optimal at high $\tsnr$s. We note from Fig.
\ref{fig:overAF1} that cooperation starts being useful as the
source-relay channel variance $\sigma^2_{sr}$ increases. Similar
results are also observed in Fig \ref{fig:overDF}. Hence, the
source-relay channel quality is one of the key factors in
determining the usefulness of cooperation in the high $\tsnr$
regime.

\begin{figure}
\begin{center}
\includegraphics[width = \figsize\textwidth]{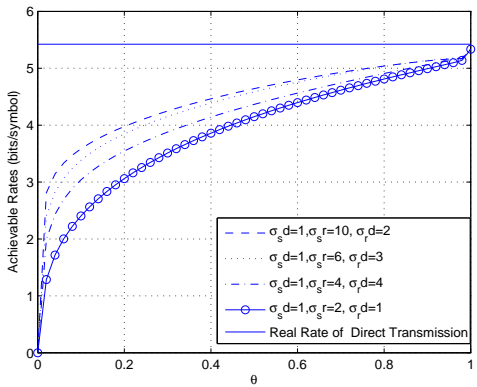}
\caption{ AF achievable rate vs. $\theta$. $P =100$. }
\label{fig:overaf}
\end{center}
\end{figure}

\begin{figure}
\begin{center}
\includegraphics[width = \figsize\textwidth]{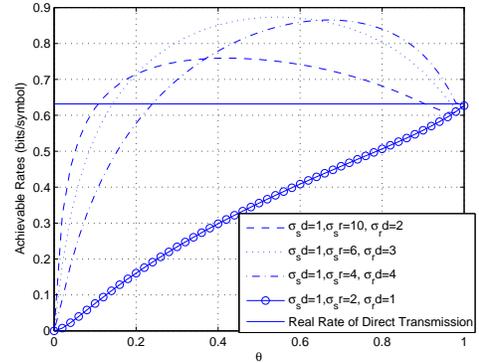}
\caption{ AF achievable rate vs. $\theta$. $P =1$. }.
\label{fig:AFoverlow}
\end{center}
\end{figure}

\begin{figure}
\begin{center}
\includegraphics[width = \figsize\textwidth]{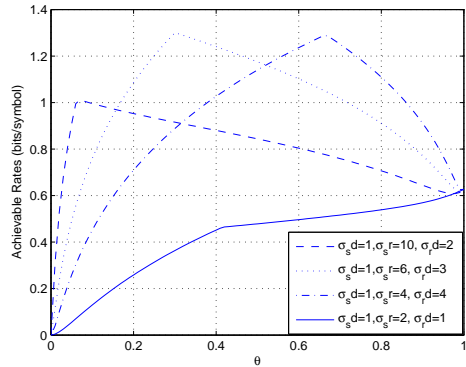}
\caption{ DF with repetition coding rate vs. $\theta$. $P =1$.
Direct transmission rate is the same as in Fig. \ref{fig:AFoverlow}.
} \label{fig:DFRoverlow}
\end{center}
\end{figure}

We now consider the special case of $\alpha=0.5$ which provides the
maximum
of degree of cooperation. 
%
%
In certain cases, source and relay are subject to a total power
constraint. Here, we introduce the power allocation coefficient
$\theta$, and total power constraint $P$. $P_s$ and $P_r$ have the
following relations: $P_s=\theta P$, $P_r=(1-\theta)P$, and
$P_s+P_r\leq P$. Next, we investigate how different values of
$\theta$, and hence different power allocation strategies, affect
the achievable rates. 
Figures \ref{fig:overaf}-\ref{fig:DFRoverlow} plot the achievable
rates as a function of $\theta$ for AF and DF. In the figures, we
have assumed that $N_0=1, \delta_s=0.1, \delta_r=0.1$\footnote{Note
that we have also obtained numerical results on optimal training
power allocations $\delta_s$ and $\delta_r$. These results are
omitted due to lack of space.}. Note that the rates for $\theta=1$
do not exactly correspond to the rates of direct transmission in
which no time is used for training for relay channels. Therefore,
for fair comparison, we also provide the direct transmission rates
in the figures.
In Fig. \ref{fig:overaf} where $P = 100$, we observe that direct
transmission without the relay is superior in this high $\tsnr$
case. On the other hand, we see different results when we turn our
attention to the low-$\tsnr$ regime. Figs. \ref{fig:AFoverlow} and
\ref{fig:DFRoverlow} provide the achievable rates of AF and DF,
respectively, when $P=1$. We observe in these cases that relaying
increases the rates and hence cooperation is useful unless the
source-relay and relay-destination channel qualities are comparable
to that of the source-destination channel.

\begin{figure}
\begin{center}
\includegraphics[width = \figsize\textwidth]{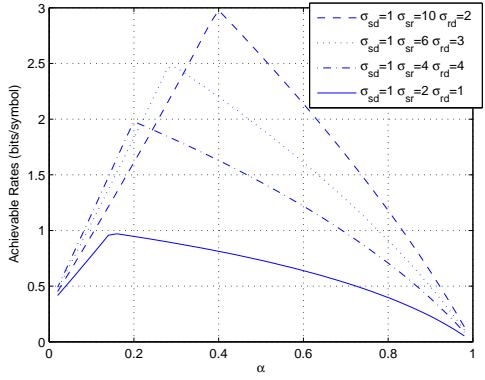}
\caption{  DF parallel coding achievable rate  vs. $\alpha$ when
$P_s=P_r=0.5, \delta_s=\delta_r=0.1$} \label{fig:overDFP}
\end{center}
\end{figure}

Finally, in Fig. \ref{fig:overDFP}, we plot the achievable rates of
DF parallel channel coding, derived in Theorem \ref{prop:asympcap}.
We can see from the figure that the best performance is obtained
when the source-relay channel quality is high (i.e., when
$\sigma_{sd} = 1, \sigma_{sr} = 10, \sigma_{rd} = 2$). Additionally,
we observe that as the source-relay channel improves, more resources
need to be allocated to the relay to achieve the best performance.
We note that significant improvements with respect to direct
transmission (i.e., $\alpha \to 0$) are obtained. Finally, we can
see that when compared to AF and DF with repetition coding, DF with
parallel channel coding achieves higher rates. On the other hand, AF
and DF with repetition coding have implementation advantages.
\section{energy efficiency}
Our analysis has shown that cooperative relaying is generally
beneficial in the low-power regime, resulting in improved achievable
rates when compared to direct transmission. In this section, we
provide an energy efficiency perspective. The least amount of energy
required to send one information bit reliably is given
by\footnote{Note that $\frac{E_b}{N_0}$ is the bit energy normalized
by the noise power spectral level $N_0$.} $\frac{E_b}{N_0} =
\frac{\tsnr}{C(\tsnr)}$ where $C(\tsnr)$ is the channel capacity in
bits/symbol. In our setting, the bit energy values are given by
$\frac{E_{b,U}}{N_0} = \frac{\tsnr}{I_{low}(\tsnr)}$.
$\frac{E_{b,U}}{N_0}$ provides the least amount of normalized bit
energy values in the worst-case scenario and also serves as an upper
bound on the achievable bit energy levels of the channel. We note
that we define the signal-to-noise ratio as $\tsnr=P/N_0$ where $P$
is the total power in the system. The next result provides the
asymptotic behavior of the bit energy as $\tsnr$ decreases to zero.
\begin{lemma:bitenergylow} \label{lemma:bitenergylow}
The normalized bit energy  in all transmission schemes grows without
bound as the signal-to-noise ratio decreases to zero, i.e.,
\begin{gather}
\left.\frac{E_{b,U}}{N_0}\right|_{I_{low} = 0} = \lim_{\tsnr \to 0}
\frac{\tsnr}{I_{low}(\tsnr)} = \frac{1}{\dot{I}_{low}(0)} = \infty.
\end{gather}
\end{lemma:bitenergylow}
The result is shown easily by proving in all relaying protocols that
$\dot{I}_{low}(0)=0$. Theorem \ref{lemma:bitenergylow} indicates
that it is extremely energy-inefficient to operate at very low
$\tsnr$ values. In general, it is not easy to identify the most
energy-efficient operating points analytically. Therefore, we resort
to numerical analysis. We choose the following numerical values for
the following parameters: $\delta_s = \delta_r = 0.1$,
$\sigma_{sd}=1$, $\sigma_{sr}=4$, $\sigma_{rd}= 4$, $\alpha = 0.5$,
and $\theta=0.6$. Fig. {\ref{fig:overAF} plots the bit energy curves
as a function of SNR for different values of $m$ in the AF case. We
can see from the figure that the minimum bit energy, which is
achieved at a nonzero value of SNR, decreases with increasing $m$
and is achieved at a lower $\tsnr$ value. Fig. \ref{fig:enper} shows
the minimum bit energy for different relaying schemes with
overlapped or non-overlapped transmission techniques. In
non-overlapped transmission, source becomes silent while the relay
transmits. The achievable rates of non-overlapped DF with parallel
coding are provided in Theorem \ref{prop:asympcap}. The
non-overlapped AF and DF with repetition coding are considered in
\cite{Zhagur}. In overlapped transmission, source continues its
transmission as the relay transmits. The rates for overlapped AF and
DF with repetition coding are given in Theorems
\ref{prop:overlapped} and \ref{prop:DFRover}. We observe in Fig.
\ref{fig:enper} that the minimum bit energy decreases with
increasing $m$ in all cases . We realize that DF is in general much
more energy-efficient than AF. Moreover, we note that surprisingly
employing non-overlapped rather than overlapped transmission
improves the energy efficiency.
\begin{figure}
\begin{center}
\includegraphics[width = \figsize\textwidth]{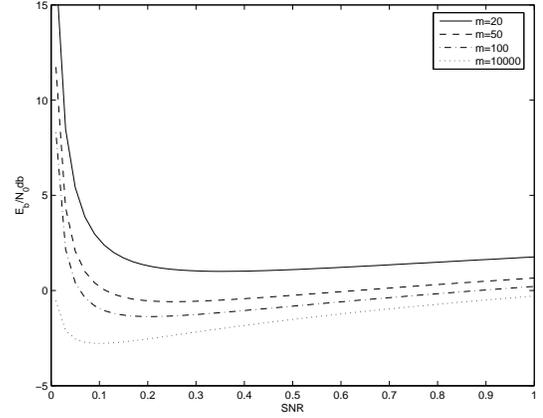}
\caption{  AF $E_{b,U}/N_0$ vs. $\tsnr$} \label{fig:overAF}
\end{center}
\end{figure}

\begin{figure}
\begin{center}
\includegraphics[width = \figsize\textwidth]{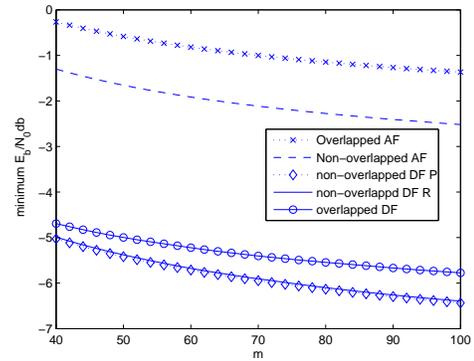}
\caption{ $E_{b,U}/N_0$ vs. $m$ for different transmission scheme. P
denotes parallel channel coding while R denotes repetition channel
coding.} \label{fig:enper}
\end{center}
\end{figure}

\end{document}